\documentclass[conference]{IEEEtran}
\IEEEoverridecommandlockouts
\usepackage{cite}
\usepackage{amsmath,amssymb,amsfonts}
\usepackage{algorithmic}
\usepackage{graphicx}
\usepackage{textcomp}
\usepackage{xcolor}
\def\BibTeX{{\rm B\kern-.05em{\sc i\kern-.025em b}\kern-.08em
    T\kern-.1667em\lower.7ex\hbox{E}\kern-.125emX}}
\begin{document}

\title{Residual Motion Compensation in \\ Automotive MIMO SAR Imaging
}

\author{\IEEEauthorblockN{Marco Manzoni}
\IEEEauthorblockA{\textit{DEIB} \\
\textit{Politecnico di Milano}\\
Milano, Italy \\
marco.manzoni@polimi.it}
\and

\IEEEauthorblockN{Marco Rizzi}
\IEEEauthorblockA{\textit{DEIB} \\
\textit{Politecnico di Milano}\\
Milano, Italy \\
marco.rizzi@polimi.it}
\and

\IEEEauthorblockN{Stefano Tebadlini}
\IEEEauthorblockA{\textit{DEIB} \\
\textit{Politecnico di Milano}\\
Milano, Italy \\
stefano.tebaldini@polimi.it}
\and

\IEEEauthorblockN{Andrea Virgilio Monti-Guarnieri}
\IEEEauthorblockA{\textit{DEIB} \\
\textit{Politecnico di Milano}\\
Milano, Italy \\
andrea.montiguarnieri@polimi.it}
\and

\IEEEauthorblockN{Claudio Maria Prati}
\IEEEauthorblockA{\textit{DEIB} \\
\textit{Politecnico di Milano}\\
Milano, Italy \\
claudio.prati@polimi.it}
\and

\IEEEauthorblockN{Dario Tagliaferri}
\IEEEauthorblockA{\textit{DEIB} \\
\textit{Politecnico di Milano}\\
Milano, Italy \\
dario.tagliaferri@polimi.it}
\and

\IEEEauthorblockN{Monica Nicoli}
\IEEEauthorblockA{\textit{DIG} \\
\textit{Politecnico di Milano}\\
Milano, Italy \\
monica.nicoli@polimi.it}
\and

\IEEEauthorblockN{Ivan Russo}
\IEEEauthorblockA{\textit{Huawei Milan Research Center} \\
Huawei Technologies Italia S.r.l.\\
Segrate, Italy \\
ivan.russo@huawei.com}
\and

\IEEEauthorblockN{Christian Mazzucco}
\IEEEauthorblockA{\textit{Huawei Milan Research Center} \\
Huawei Technologies Italia S.r.l.\\
Segrate, Italy \\
christian.mazzucco@huawei.com}
\and

\IEEEauthorblockN{Sergi Duque Biarge}
\IEEEauthorblockA{\textit{Huawei German Research Center}\\
Huawei Technologies Duesseldorf GmbH \\
Munich, Germany \\
sergio.duque.biarge@huawei.com}
\and

\IEEEauthorblockN{Umberto Spagnolini}
\IEEEauthorblockA{\textit{DEIB} \\
\textit{Politecnico di Milano}\\
Milano, Italy \\
umberto.spagnolini@polimi.it}
}
\maketitle
\begin{abstract}
This paper deals with the analysis, estimation, and compensation of trajectory errors in automotive-based Synthetic Aperture Radar (SAR) systems. First of all, we define the geometry of the acquisition and the model of the received signal. We then proceed by analytically evaluating the effect of an error in the vehicle's trajectory. Based on the derived model, we introduce a motion compensation (MoCo) procedure capable of estimating and compensating constant velocity motion errors leading to a well-focused and well-localized SAR image.\\
The procedure is validated using real data gathered by a 77 GHz automotive SAR with MIMO capabilities.
\end{abstract}


\section{Introduction}
Fully-autonomous vehicles will be equipped with a huge set of different sensors, such as cameras, LiDARs, radars, acoustic, etc. However, the camera/LiDAR sensitivity to adverse weather conditions challenges their usage for safety critical applications. In this regard, automotive-legacy multiple-input multiple-output (MIMO) radars working in W-band ($76-81$ GHz~\cite{Hasch2012}) are widely employed to obtain measures of radial distance, velocity and angular position of remote targets, but are characterized by a medium-to-low angular resolution. 
Significant effort was spent in recent works to increase the accuracy of environmental perception by leveraging Synthetic Aperture Radar (SAR) techniques \cite{Feger2017_experimentalSAR77GHz, iqbal_imaging_2021}.\\
In \cite{feger_experimental_2017}, a 77 GHz radar with 1 GHz of bandwidth was installed on the rooftop of a car to obtain images with resolution as small as 15 cm, while in \cite{wang_auxiliary_2019} SAR images were used to search for free parking areas. A 300 GHz implementation of a car-borne SAR is presented in \cite{stanko_millimeter_2016}.
All these implementations, however, are not fully reflecting a real scenario: either the radar is mounted on a rail and not on a car, or expensive and non-commercial positioning systems are used, or extremely low car velocities are considered. A real scenario is presented in \cite{tagliaferri_navigation-aided_2021}, where a 77 GHz MIMO radar has been mounted on a car and an automotive-grade multi-sensor integrated navigation solution is used to enable cm-accurate environment mapping.

This paper tackles the problem of residual motion estimation. In order to obtain a well-focused image, the SAR processor needs to know the exact position of the Antenna Phase Centers (APC) at any given time. 
This requirement makes it mandatory to use advanced automotive-grade navigation solutions, e.g., fusing Global Navigation Satellite System (GNSS), Inertial Measurement Units (IMU) and other sensors' data, but the result could still be not sufficiently accurate for automotive SAR purposes \cite{fornaro_motion_2005, azouz_motion_2014}.
The solution proposed herein exploits a set of low resolution range/azimuth images generated by a MIMO radar installed on the vehicle. From these images, the motion error is estimated and the results are used to correct the data before SAR focusing.


\section{Acquisition geometry and signal model}
\begin{figure}[!t]
    \centering
    \includegraphics[width=0.8\columnwidth]{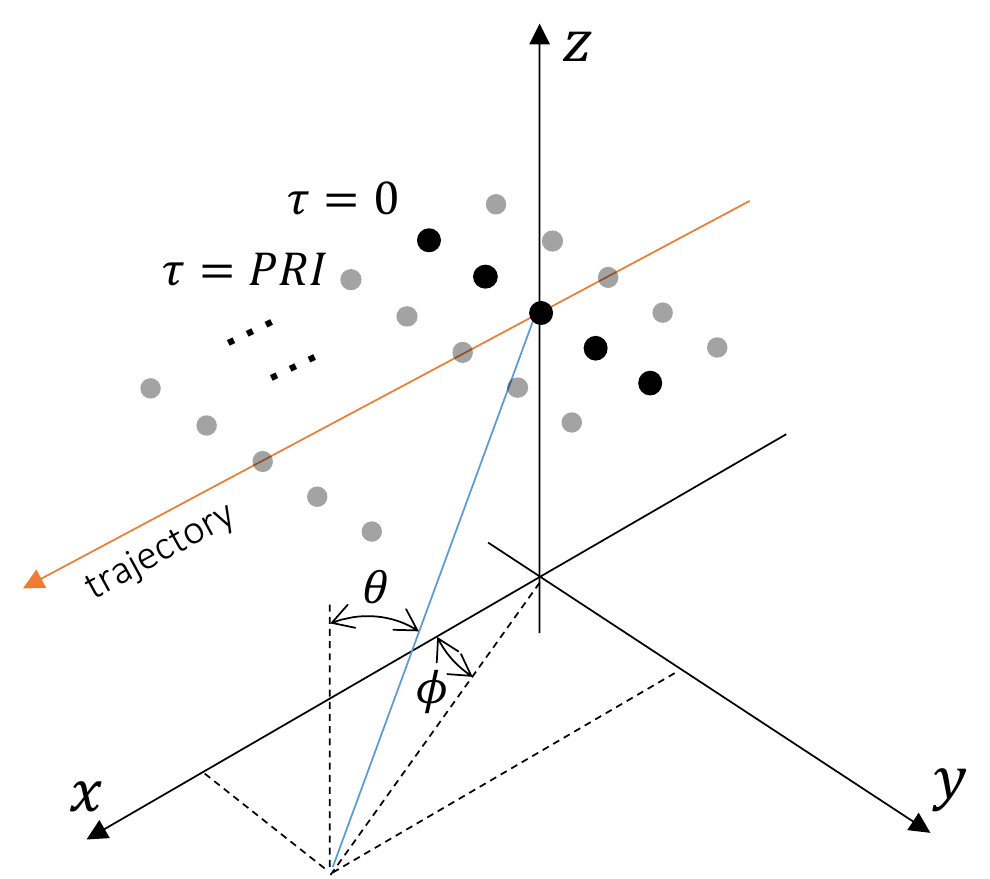}
    \caption{SAR intensity image without employing any MoCO procedure.}
    \label{fig:geometry_MIMO}
\end{figure}

We consider a MIMO radar with $N$ Virtual Phase Centers (VPC) displaced along $y$, which is the direction orthogonal to the motion. Therefore, the radar is mounted in a forward looking configuration, i.e., with the boresight in the direction of motion $x$, Fig. \ref{fig:geometry_MIMO}. 
We start by writing the model of the baseband and range compressed received signal from a target in the Field of View (FoV):
\begin{equation}
\begin{split}
        s_{rc}(r,n,\tau; \mathbf{x}_0) = &\alpha\, \mathrm{sinc}\left[\frac{r-r(n,\tau ; \mathbf{x}_0)}{\rho_r}\right] \times \\
        & \times \mathrm{exp}\left\{-j\frac{4\pi}{\lambda}r(n,\tau ; \mathbf{x}_0)\right\}
\end{split}
\end{equation}
where $\alpha$ is a complex factor representing the propagation losses and the backscattering of the target, $r$ is the range, $n$ is the index of the virtual phase center, $\tau$ is the slow-time, $\mathbf{x}_0$ is the vector representing the 3D position of the target and $r(n,\tau ; \mathbf{x}_0)$ is the distance from the $n$-th VPC to the target at the slow time $\tau$.

The distance $r(n,\tau ; \mathbf{x}_0)$ can be written under the plane wave approximation:
\begin{equation}
    r(n,\tau ; \mathbf{x_0}) \approx r(\tau; \mathbf{x}_0) - n\Delta y\, \mathrm{sin}(\theta_0)\mathrm{sin}(\phi_0)
\end{equation}
where $r(\tau; \mathbf{x}_0)$ is the distance from the center of the virtual array to the target, $\Delta y$ is the spacing between VPCs, $\theta_0$ and $\phi_0$ are the incidence angle and the azimuth angle, respectively.
The distance $r(\tau; \mathbf{x}_0)$ can be also approximated, considering a vehicle moving at a constant velocity $\mathbf{v} = [v_x,v_y,v_z]^\mathrm{T}$ along the synthetic aperture:
\begin{equation}
\label{eq:range_equation}
\begin{split}
        r(\tau; \mathbf{x}_0) &\approx r_0 - \left(\frac{x_0}{r_0}v_x + \frac{y_0}{r_0}v_y + \frac{z_0}{r_0}v_z\right)\tau = \\
        &= r_0 - (\mathbf{u}(\mathbf{x}_0) \cdot \mathbf{v})\tau
\end{split}
\end{equation}
where $r_0 = r(0; \mathbf{x}_0)$ is the distance from the center of the synthetic aperture (supposed at $\tau=0$) to the target in $\mathbf{x}_0$, "$\cdot$" denotes the dot product and $\mathbf{u}(\mathbf{x}_0)$ is the unit vector pointing from the center of the synthetic aperture to the target. Following the geometry of Figure \ref{fig:geometry_MIMO} we can also write $\mathbf{u}(\mathbf{x}_0)$ as:
\begin{equation}
    \mathbf{u}(\mathbf{x}_0) = [\mathrm{sin}(\theta_0)\mathrm{cos}(\phi_0), \, \mathrm{sin}(\theta_0)\mathrm{sin}(\phi_0), \, \mathrm{cos}(\theta_0)]^{\mathrm{T}}.
\end{equation}
While the car moves forward, the radar transmits a pulse every Pulse Repetition Interval (PRI). A synthetic aperture is formed by coherently processing a set of $M$ pulses.

Once the data have been acquired, the SAR focusing is rather straightforward. First of all, $M$ low resolution images are generated by Time Domain Back Projection (TDBP). Each image is formed by backprojecting all the $N$ range compressed signals at time $\tau$:
\begin{equation}
\label{eq:TDBP_MIMO}
    I_m(\tau ; \mathbf{x}) \hspace{-0.1cm}=\hspace{-0.1cm} \sum_{n=1}^N s_{rc}[r(n,\tau ; \mathbf{x}),n,\tau;\mathbf{x}_0]\mathrm{exp}\left\{j\frac{4\pi}{\lambda}r(n,\tau ; \mathbf{x})\right\}.
\end{equation}
The resolution of $I_m(\tau ; \mathbf{x})$ is dictated by the length of the virtual array, and it is maximum at radar's boresight (i.e., in front of the car). The second step is the SAR processing, consisting of a simple coherent sum of all the low-resolution images in the synthetic aperture time $T$:
\begin{equation}
\label{eq:coherent_summation}
       I(\mathbf{x}) = \sum_{\tau \in T} I_m(\tau ; \mathbf{x}).  
\end{equation}
The resolution of the SAR image is ruled by $T$ and it is higher in the direction orthogonal to the motion (i.e.; at the right/left of the car).
In order to compute $r(n,\tau ; \mathbf{x})$ in \eqref{eq:TDBP_MIMO}, the knowledge of the  position of each VPC for each of the $M$ pulses must be very accurate. In the next section, we discuss the error generated by a wrong knowledge of the vehicle's trajectory.

\section{Motion Error Analysis}\label{sec:motion_error_analysis}
Let us assume that the knowledge of the vehicle motion is effected by errors. In other words, the on-board navigation unit provides an erroneous estimate of the vehicle velocity:
\begin{equation}
    \mathbf{v}_{nav} = \mathbf{v} + \Delta \mathbf{v}
\end{equation}
where $\Delta \mathbf{v} = [\Delta v_x,\Delta v_y,\Delta v_z ]^\mathrm{T}$ is the velocity error. Notice that we are assuming a constant velocity error: this is reasonable for short synthetic aperture.

Eq.~\eqref{eq:range_equation} shows that the range is linear with velocity, therefore an error in the latter will be linearly transferred to the former: 
\begin{equation}
    \widetilde{r}(n,\tau; \mathbf{x}) = r(n,\tau; \mathbf{x}) - (\mathbf{u}(\mathbf{x}) \cdot \Delta \mathbf{v}) \tau 
\end{equation}
Where $\widetilde{r}(n,\tau; \mathbf{x})$ is the erroneous distance generated by the error $\Delta \mathbf{v}$. The direct consequence is that, when the backprojection of Eq. \eqref{eq:TDBP_MIMO} is performed, a linear residual phase term arises:
\begin{equation}
\label{eq:TDBP_MIMO_error}
\begin{split}
        \widetilde{I}_m(\tau ; \mathbf{x}) &=\hspace{-0.1cm} \sum_{n=1}^N s_{rc}[\widetilde{r}(n,\tau ; \mathbf{x}),n,\tau;\mathbf{x}_0]\mathrm{exp}\left\{j\frac{4\pi}{\lambda}\widetilde{r}(n,\tau ; \mathbf{x})\right\} \\
        &= I_m(\tau ; \mathbf{x}) \times \mathrm{exp}\left\{-j\frac{4\pi}{\lambda}\mathbf{u}(\mathbf{x}) \cdot \Delta \mathbf{v} \tau \right\}.
\end{split}
\end{equation}
The phase error is linear with $\tau$ and the slope is proportional to the \textit{radial} residual velocity error $\mathbf{u}(\mathbf{x}) \cdot \Delta \mathbf{v}$ (i.e.; the velocity error projected in the looking direction). 
If this phase term is not compensated, the SAR processing represented by the coherent summation of \eqref{eq:coherent_summation} does not lead to a well-focused image.

The degradation of the SAR image is particularly strong in those areas of the FoV pointed by the vector $\Delta \mathbf{v}$: in these locations, in fact, the residual radial velocity is at its maximum leading to a severe defocusing.
It is easy to show that a trajectory error is not the only one that generates a residual linear phase: an error in the focusing height generates the same effect. In the following, we analyze the problem and assess the impact in an automotive scenario.

\subsection{Error in the focusing height}
In Eq. \eqref{eq:TDBP_MIMO}, the 3D coordinate $\mathbf{x}$ of every pixel of the backprojection grid can be chosen arbitrarily. It is common practice to choose a fixed focusing height. In particular, we can decide to focus the image at a fixed height $q$, thus $\mathbf{x} = [x,y,z=q]^\mathrm{T}$. In a real scenario, however, the height of the target is not known and it is possibly different from $q$. If the target is not truly at that height, another residual phase term (distance error) arises. The expression of such residual distance can be easily derived by considering that an error in the knowledge of the target's height $q$ is equal to an error in the knowledge of the incidence angle $\theta$, thus:
\begin{equation}
\label{eq:residual_phase_elevation}
\begin{split}
        \Delta r^q(\tau; \mathbf{x}) &= \frac{\partial r(\tau; \mathbf{x})}{\partial \theta}\Delta \theta = \tau \left(\mathbf{u}'(\mathbf{x})\cdot \mathbf{v} \right)  \Delta \theta 
\end{split}
\end{equation}
where $\mathbf{u}'(\mathbf{x})$ is the derivative of the vector $\mathbf{u}(\mathbf{x})$ with respect to $\theta$ and $\Delta \theta = \Delta q/(r\sin\theta)$, $\Delta q$ being the mismatch between the focusing height and the real elevation of the target.

Notice that \eqref{eq:residual_phase_elevation} is again linear with respect to $\tau$: an error in the velocity and an error in the focusing height manifest as a linear phase in time. The understanding of which of the two effects is more relevant is of utmost importance in automotive systems. Let's consider a constant nominal velocity along $x$ and the azimuth position that maximizes the error due to elevation mismatch ($\phi = 0$, in the direction of motion). We have:
\begin{equation}
    \Delta r^q(\tau; \mathbf{x}) = \frac{v_x \tau \Delta q}{r \tan\theta}.
\end{equation}
For a front looking radar, where the the radar is installed on the frontal bumper, the elevation angle is $\theta \approx 90$ deg for all the FoV, except for the very close range and the error in the height can be in the order of few meters depending on the radar's antennae aperture.

Considering a vehicle travelling at a nominal velocity of $v_x = 15$ m/s and a target at slant range $R = 20$ m, characterized by an incidence angle $\theta = 87$ deg, for a height error $\Delta q = 1$ m and an operational wavelength $\lambda = 4$ mm, we have a residual Doppler frequency of $19$ Hz. The same residual Doppler frequency is induced by a velocity error along the direction of motion of $\Delta v_x = 3.9$ cm/s, much less than the typical velocity accuracy provided by an automotive-grade navigation system, that ranges from $10$ to $30$ cm/s. Therefore, the effect of a mismatch between the real target position in elevation and the focusing plane is much less relevant than a residual velocity. To mitigate this effect, it is good practice to use Ground Control Points (GCP, see Section \ref{sec:moco}) with high incidence angle. Nevertheless, if the employed radar is equipped with real of virtual antennas displace along the vertical direction, it is be possible to jointly estimate elevation and residual velocity.

\section{Motion Error Compensation}
\label{sec:moco} 

\begin{figure}[!t]
\centering
\includegraphics[width=0.8\linewidth]{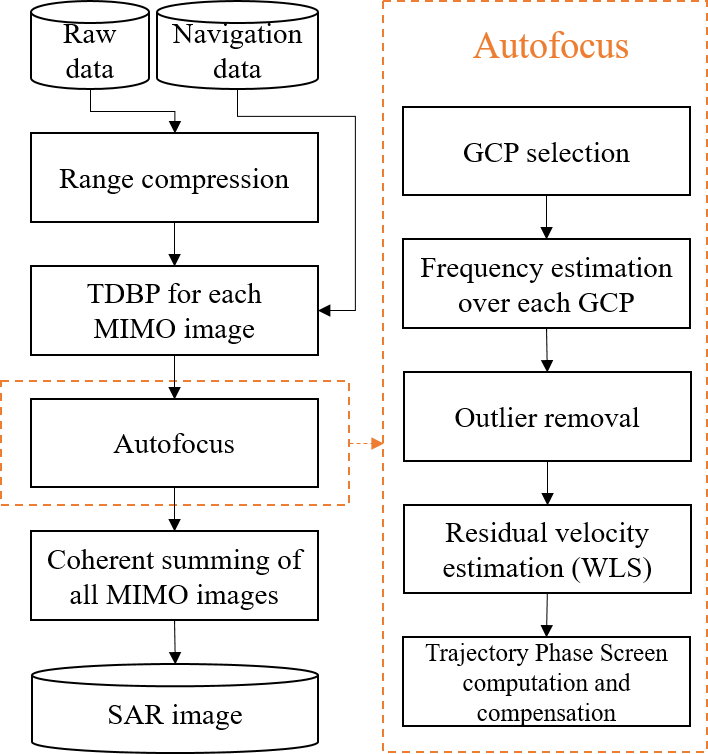}
\caption{(left) The trajectory is perfectly know during the focusing routine. (right) the trajectory contains a linear error in the $x$ direction. The scene is rotated as predicted by the theory.}
\label{fig:block_diagram}
\end{figure}

In this section, we propose a Motion Compensation (MoCo) technique to estimate and compensate trajectory errors starting from a set of low-resolution images provided by the MIMO array. Figure \ref{fig:block_diagram} shows the complete SAR data processing and autofocus workflow.

\subsection{From raw data to a stack of low-res images}

We assume the vehicle equipped with a MIMO FMCW array able to generate an equivalent virtual monostatic array of $N$ elements. The signal received by each of the $N$ channels is first range compressed and then backprojected with \eqref{eq:TDBP_MIMO}, generating the $M$ low resolution images $I_m(\tau; \mathbf{x})$, $m=1,...,M$. Each image is formed by back-projecting in the Field of View (FoV) the $N$ signals received at each VPC of the array at the time instant $\tau$. If the TDBP is performed over a common grid (FoV) for all the slow time instants, all the images are already compensated for range migration.

\subsection{Autofocus} 

The input to the MoCo procedure is the set of $M$ low resolution images $I_m(\tau; \mathbf{x})$, $m=1,...,M$. As already discussed in Section \ref{sec:motion_error_analysis}, the presence of a constant velocity error $\Delta \mathbf{v}$ leads to a linear residual range (or phase) after TDBP.
The autofocus procedure starts from \eqref{eq:TDBP_MIMO_error}: the exponential term can be regarded as a complex sinusoid with angular frequency:
\begin{equation}
\label{eq:frequency_residual}
   \Delta \omega(\mathbf{x}) =  \mathbf{k}(\mathbf{x}) \cdot \Delta\mathbf{v},
\end{equation}
where $\mathbf{k}(\mathbf{x}) = (4\pi/\lambda)\mathbf{u}(\mathbf{x})$ in the wavevector in the direction of $\mathbf{x}$. Equation \eqref{eq:frequency_residual} provides a single equation for three unknowns (the three components of the velocity error). Exploiting the linear array, providing low-resolution images at each slow time, it is now sufficient to detect a few stable Ground Control Points (GCP) in the scene to have an overdetermined linear system of equations. If we consider a total of $V$ GCP we can write:
\begin{equation}
\label{eq:linear_system}
    \underbrace{\begin{bmatrix}
        \Delta \omega(\mathbf{x}_0)  \\
        \Delta \omega(\mathbf{x}_1)  \\
        \Delta \omega(\mathbf{x}_2)  \\
        \vdots \\
        \Delta \omega(\mathbf{x}_{V-1}) 
    \end{bmatrix}}_{\Delta \mathbf{\Omega}}
    = 
    \underbrace{\begin{bmatrix}
    k_x^0 & k_y^0 & k_z^0\\
    k_x^1 & k_y^1 & k_z^1\\
    k_x^2 & k_y^2 & k_z^2\\
    \vdots & \vdots & \vdots\\
    k_x^{V-1} & k_y^{V-1} & k_z^{V-1}
    \end{bmatrix}}_{\mathbf{K}}
    \underbrace{\begin{bmatrix}
    \Delta v_x \\
    \Delta v_y \\
    \Delta v_z
    \end{bmatrix}}_{\Delta \mathbf{v}}
    +
    \underbrace{\begin{bmatrix}
        n_0  \\
        n_1  \\
        n_2  \\
        \vdots \\
        n_{V-1}
    \end{bmatrix}}_{\mathbf{n}}
\end{equation}
\begin{figure}[!t]
    \centering
    \includegraphics[width=0.8\columnwidth]{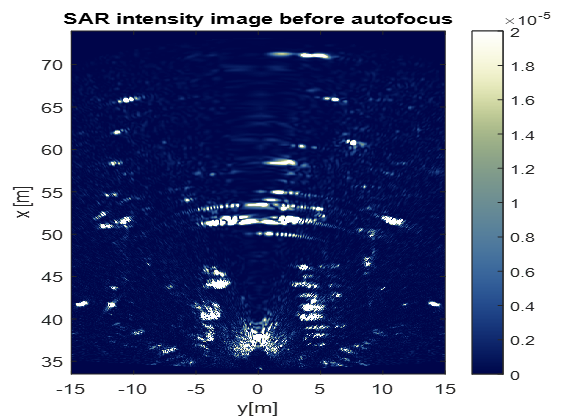}
    \caption{SAR intensity image without employing any MoCO procedure.}
    \label{fig:no_autofocus}
\end{figure}
where $\mathbf{n}$ is the vector representing the noise on the estimates of the residual frequencies $\Delta \mathbf{\Omega}$.
The process of selection of stable GCP can be easily performed by looking at the amplitude statistics of the scene. In particular we compute the \textit{incoherent} average (i.e. the average of the amplitudes) of all the $M$ low resolution images and select only the brightest targets. Since the linear system \eqref{eq:linear_system} has only three unknowns, it is sufficient to have a few anchors ($20-30$) to obtain a reliable estimate.
For every GCP, a frequency estimation is performed through a Fast Fourier Transform (FFT) and the position in of the peak in the frequency domain is extracted to form the array $\Delta \mathbf{\Omega}$ in \eqref{eq:linear_system}.

It is possible that a moving target (bike, vehicle, pedestrian, etc.) is detected as a GCP, preventing a correct residual motion estimation. It is mandatory to discard the outliers before the inversion of \eqref{eq:linear_system}: this step can be performed by imposing a threshold on the maximum frequency that is possible to associate to a given GCP. The value of the threshold can be derived from the accuracy of the navigation data: if the the nominal accuracy is, for instance, $20$ cm/s, it is unlikely to find a GCP with a residual Doppler frequency much higher than the one corresponding to $20$ cm/s.
The linear system is inverted using Weighted Least Square (WLS):
\begin{equation}
\label{eq:residual_velocities}
    \widehat{\Delta}\mathbf{v} = (\mathbf{K}^T\mathbf{W}\mathbf{K})^{-1}\mathbf{K}^T\mathbf{W}\Delta \mathbf{\Omega}
\end{equation}
\begin{figure}[!t]
    \centering
    \includegraphics[width=0.8\columnwidth]{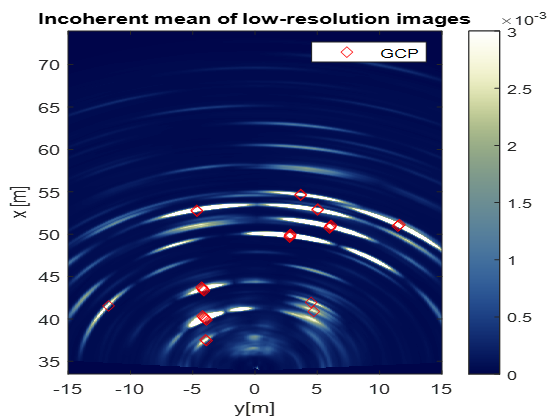}
    \caption{Incoherent mean of all the low resolution images. The selected GCP are depicted with red dots.}
    \label{fig:mimo_anchors}
\end{figure}
where $\mathbf{W}$ is a matrix of weights. Each GCP can be weighted according to some specific figure of merit such as the amplitude of the GCP or the prominence of the peak in the frequency domain.

Notice that \eqref{eq:linear_system} can provide an unstable solution. In a typical automotive environment, the radar is installed close to the road, thus $\theta \approx 90$ deg for every GCP. This configuration prevents any reliable estimate of $\Delta v_z$. The solution can be to avoid the estimation of the residual velocity in the vertical direction by removing the last column of $\mathbf{K}$ and the last row of $\Delta\mathbf{v}$.
Once the residual velocities has been estimated, it is possible to compute the residual frequencies with \eqref{eq:linear_system} (forward problem) for each pixel $\mathbf{x}$ in the scene and for each $\tau$ (i.e. each one of the $M$ low resolution MIMO images). This leads to a set of estimated phase screens:
\begin{equation}
    \widehat{\Delta}\psi(\mathbf{x},\tau) = -\left(\mathbf{k}(\mathbf{x}) \cdot \widehat{\Delta}\mathbf{v}\right)\tau
\end{equation}
used to compensate each low resolution image as
\begin{equation}
    I_m^c(\mathbf{x},\tau) = I_m(\mathbf{x},\tau)\exp\left\{-j\widehat{\Delta}\psi(\mathbf{x},\tau)\right\}.
\end{equation}
The residual velocities estimated in \eqref{eq:residual_velocities} can be used also to improve the ego-motion estimation of the vehicle. In this case it is sufficient to integrate the residual velocities to obtain the residual trajectory and then compensate the error in the original trajectory provided by the navigation unit.\\
Once the MoCo procedure has been performed, the coherent sum of all the phase-compensated low resolution images provides the correct high resolution SAR image
\begin{equation}
    I(\mathbf{x}) = \sum_{\tau \in T}I_m^c(\mathbf{x},\tau)
\end{equation}
\section{Results with real data}
\begin{figure}[!t]
    \centering
    \includegraphics[width=1\columnwidth]{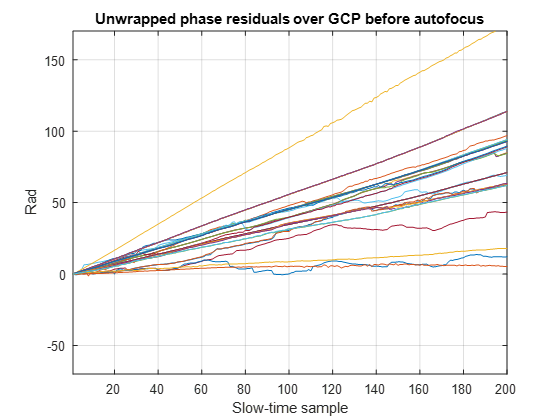}
    \caption{Unwrapped phase over the selected GCP: a residual phase is linear in slow time indicating the presence of a residual motion error}
    \label{fig:unwrapped}
\end{figure}

To validate the proposed MoCo technique, we carried out an acquisition campaign using a fully equipped vehicle. The radar system is a proprietary ScanBrick\textregistered by Aresys\textregistered and it is based on the AWR1243 single-chip $77$ and $79$ GHz FMCW Transceiver by TI. The radar is mounted in a forward looking geometry, thus the boresight of the MIMO array is pointing in the direction of motion. The VPC displaced along the direction orthogonal to motion are needed for the suppression of left/right ambiguity typical of SAR systems. In this campaign, we employed two transmitting antennas and four receiving ones leading to a virtual array of $8$ elements spaced by $\lambda/4$. The angular resolution of the low-resolution images is roughly $16$ deg. The transmitted signal has $3$ GHz of bandwidth leading to a range resolution of $5$ cm. The car equipment is complemented by both on-board (integrated by car manufacturer) and dedicated (mounted for the experiments) navigation sensors such as three internal Inertial Measurement Units, four wheel encoders, a steering angle sensors and a GNSS module. The navigation data are fused with an Unscented Kalman Filter (UKF) as indicated in \cite{tagliaferri_navigation-aided_2021} to provide the initial position and velocity estimates as input to the MoCo procedure.


We selected a portion of the trajectory made by $M=200$ slow time samples as the synthetic aperture and we processed the dataset with and without running the MoCo procedure. The nominal speed of the vehicle in the selected synthetic aperture was $\approx 25$ km/h.

The result of the processing without MoCo is depicted in Figure \ref{fig:no_autofocus}.
It is interesting to notice how in far range the image is completely corrupted by an error on both along-track and across-track velocity provided by the navigation data. The image seems to collapse inward. Some details are still preserved and not totally defocused in near range (i.e. the bikes and the cars parked for $y=\pm 5$ m. Nevertheless, the image quality is not sufficient.

In Figure \ref{fig:mimo_anchors}, the incoherent average of the low resolution images is represented along with the detected GCP (in red). It is easy to see that the spatial resolution of the image formed with the virtual array is much lower than the one formed by coherent SAR processing, but low resolution images are not severely corrupted by trajectory error as the SAR one.
\begin{figure}[!t]
    \centering
    \includegraphics[width=0.8\columnwidth]{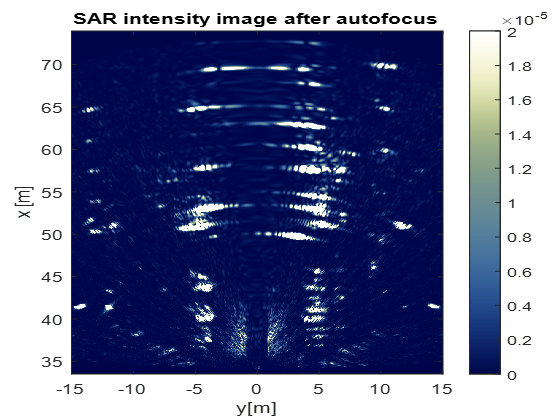}
    \caption{SAR intensity image after employing the proposed MoCO procedure.}
    \label{fig:autofocus}
\end{figure}
While the cars parked in the scene at $y = -5$ m and $x = [35,45]$ m are tilted inwards in Figure \ref{fig:no_autofocus}, they are straight (but with lower resolution) in Figure \ref{fig:mimo_anchors}. This is an expected behavior: the cars are parked straight, but the coherent processing with a wrong trajectory generates a significant displacement in the reconstructed scene.

In Figure \ref{fig:unwrapped}, the unwrapped residual phase over all the selected GCP is depicted. It is evident that the phase is roughly linear for every detected GCP and the slope is proportional to the residual radial velocity. 
\begin{figure*}[!t]
    \centering
    \includegraphics[width=0.8\textwidth]{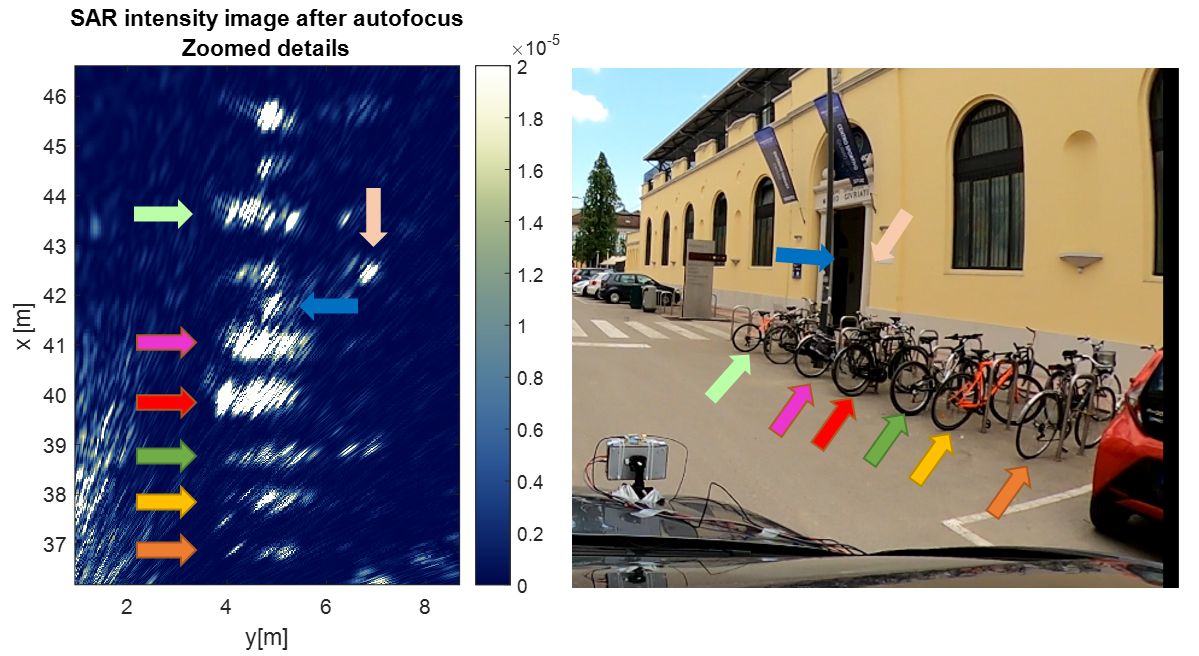}
    \caption{(left) Some details in the SAR image. (right) optical image with the same details highlighted by arrows. Notice on the lower left of the image the radar mounted on the vehicle.}
    \label{fig:details_1}
\end{figure*}
The residual frequency over each GCP is estimated and the inverse problem is solved leading to an estimate of the residual velocities and accuracy on the estimation of such velocities detailed in Table \ref{table:residual_velocities}. The residual velocities found are within the accuracy bound of the sensors for the navigation \cite{tagliaferri_navigation-aided_2021}.
\begin{table}[!b]
\renewcommand{\arraystretch}{1.3}
\caption{Residual velocities and accuracy}
\label{table:residual_velocities}
\centering
\begin{tabular}{|c||c||c|}
\hline
\textbf{Parameter} & \textbf{Value (cm/s)} & \textbf{Accuracy (cm/s)}\\
\hline
$\Delta v_x$ & 26.22 & 1.48\\
$\Delta v_y$ & -1.14 & 2.93\\
\hline
\end{tabular}
\end{table}
Once the velocities are correctly estimated, each low resolution image is phase compensated and then the coherent average is taken to form the final SAR image. In Figure \ref{fig:autofocus}, the reconstructed SAR image after the application of the proposed autofocusing method is reported. Differently from Figure \ref{fig:no_autofocus}, the profile given by the parked cars is straight also in far range, as expected. A few details are depicted in Figure \ref{fig:details_1}. On the left of the figure, a zoomed version of Figure \ref{fig:autofocus} is presented. We can distinguish the five bicycles parked at the right of the road (orange, yellow, green, red and purple arrows), the lighting pole (blue arrow), the marble column (pink arrow) and the next bicycle after the lighting pole (light green arrow). On the right, the optical image is reported for comparison.

\section{Conclusions}
In this paper we presented a residual MoCo algorithm for automotive-based MIMO SAR. The workflow starts from a set of low-resolution images gathered by a MIMO radar mounted on the vehicle in a forward looking configuration. We proceed by estimating the residual Doppler frequencies and associated residual velocities on a set of GCP in the scene. The low-resolution MIMO images are then phase-calibrated before SAR focusing using the estimated residual velocities. The entire procedure is validated using real data, obtaining correctly focused SAR images in real driving conditions. In addition, the estimated residual velocities by the MoCo procedure can be used to provide a better ego-motion accuracy.

\section*{Acknowledgment}
The research has been carried out in the framework of the Huawei-Politecnico di Milano Joint Research Lab on automotive SAR. The Authors want to acknowledge Dr. Paolo Falcone from Aresys for the cooperation and support in the data acquisition.

\bibliographystyle{IEEEtran}
\bibliography{references_zotero.bib,Bibliography.bib}


\end{document}